# Numerical method to optimize the Polar-Azimuthal Orientation of Infrared Superconducting Nanowire Single-Photon Detectors

Mária Csete,[1, 2,*] Áron Sipos,[1] Faraz Najafi,[2] Xiaolong Hu,[2] and Karl K. Berggren[2, 3]

[1]*Department of Optics and Quantum Electronics, University of Szeged,*
*Dom ter 9, Szeged, H-6720, Hungary*

[2]*Research Laboratory of Electronics, Massachusetts Institute of Technology,*
*77 Massachusetts Avenue, Cambridge, MA 02139, USA*

[3]*Currently Kavli Institute for Nanoscience, Delft University of Technology,*
*Lorentzweg 1, 2628CJ, Delft, The Netherlands*

[*]*Corresponding author: mcsete@mit.edu*

**Abstract**

A novel finite-element method for calculating the illumination-dependence of absorption in three-dimensional nanostructures is presented based on the RF module of the COMSOL software package. This method is capable of numerically determining the optical response and near-field distribution of sub-wavelength periodic structures as a function of illumination orientations specified by polar angle, $\varphi$, and azimuthal angle, $\gamma$. The method was applied to determine the illumination-angle-dependent absorptance in cavity-based superconducting-nanowire single-photon detector (SNSPD) designs. Niobium-nitride stripes based on dimensions of conventional SNSPDs and integrated with ~ quarter-wavelength hydrogensilsesquioxane-filled nano-optical cavities and covered by a thin gold film acting as a reflector were illuminated from below by p-polarized light in this study. The numerical results were compared to results from complementary transfer-matrix-method calculations on composite layers made of analogous film-stacks. This comparison helped to uncover the optical phenomena contributing to the appearance of extrema in the optical response. This paper presents an approach to optimizing the absorptance of different sensing and detecting devices via simultaneous numerical optimization of the polar and azimuthal illumination angles.

# 1. Introduction

Superconducting-nanowire single-photon detectors (SNSPD) are used for infrared photon counting due to their high quantum efficiency combined with high counting rates, low timing jitter, and low dark counts [1].

The typical dimensions used in commercially available devices are 200 nm periodicity, 50 % filling factor, and 4 nm thickness. Previous studies have shown that integration of an optical cavity and anti-reflection-coatings (ARC) can be used to overcome significant fraction of the optical losses, and to reach a device detection efficiency (*DE*) ~ 50 % [2]. Here, the device detection efficiency is the overall efficiency, normalized by the illuminated area, and thus neglects problems associated with coupling light into the active area of the detector. This device detection efficiency depends on the electronic and optical efficiency according to the relation $DE = P_R \cdot A$, where *A* refers to the absorptance of the detector and $P_R$ is the probability that an absorbed photon leads to a measurable electronic signal. This dependence implies that SNSPD detectors can be optimized optically by maximizing the absorptance in the NbN.

The absorbed light intensity depends on the orientation of the **E**-field oscillation with respect to NbN wires in case of polarized light illumination [3]. This strong polarization dependence of the detection efficiency of the NbN pattern in SNSPDs can be a disadvantage, or can be useful, depending on the application. Therefore, novel methods of understanding and controlling the sensitivity of the detector to light with arbitrary polarization are of interest.

The results of two-dimensional finite-element-method (FEM) computations performed previously by Anant et al. [3] for the case of perpendicular front-side illumination indicated higher absorptance for **E**-field oscillation parallel to the wires. The absorptance in this case was independent of the device period but did depend on the filling-factor. It was further shown that the smaller absorptance in case of **E**-field oscillation perpendicular to the wires might be enhanced by increasing the grating period and the wire width simultaneously, so as to keep the filling factor constant [3].

Another absorptance-maximization approach uses oblique illumination. One group successfully applied the transfer matrix method (TMM) to this problem, which was originally developed to determine the illumination-angle-dependent optical response of unpatterned multilayers [4]. This group predicted nearly perfect absorptance in NbN thin films of the appropriate thickness illuminated at the angle of Total Internal Reflection (TIR) by s-polarized light [5]. As the NbN nanowire arrays are similar to wire-grid polarizers, to analyze their polarization-dependent absorptance, rigorous coupled-wave analysis (RCWA) was adopted [6] from the theory of wire-grid-polarizers [5]. The two specific periodic-pattern orientations are referred to as either P-structures, when the wires are parallel to the plane of incidence, or as S-structures, when they are perpendicular to the plane of incidence [7]. In [5], the RCWA method was used to show that at the angle of TIR the absorption for s-polarized light illuminating an S-structure approached $A$ = 100 %, (similar to the result for continuous films), while the absorption for p-polarized light illuminating a P-structure vanished.

In theoretical studies, several limiting factors arise when the **E**-field oscillation is perpendicular to the wires (e.g. p-polarized illumination of S-structures), since in this case it is not straightforward to define an effective dielectric constant, and it is not possible to implement a simple impedance model, which would be correct only for polarization parallel to the wires [8].

In order to determine the optimal illumination condition for detectors composed of two-dimensional periodic patterns, the modal analysis developed to analyze noble-metal wire-grid polarizers in off-axis illumination, i.e. when the polar angle is tuned; and in arbitrary conical-mountings, i.e. when the relative orientation of the plane of light incidence and the grating-grooves is varied, would be the most appropriate method [9]. Semi-analytic procedures such as modal analysis can be enormously time consuming, as a result of the complexity of sub-wavelength periodic structures in detector designs such as SNSPDs. Thus, for rapid progress in nano-optical device analysis, it is necessary to develop numerical methods to replace these procedures. This need motivated the selection of FEM in our present work.

The purpose of our present studies was to develop a convenient numerical method for analyzing the optical near-field and absorptance of the NbN pattern illuminated by arbitrary orientations of **E**-field oscillation relative to the stripes. The novelty of our present approach is in the use of a numerical as opposed to analytic or semi-analytic investigation for integrated sub-wavelength structures for off-axis illumination and arbitrary conical-mounting. To illustrate the utility of this method, we determined the illumination-angle-dependent absorptance of NbN wires embedded in optical cavities, illuminated by p-polarized light from the substrate side, in contrast to previous work that studied only structures without a cavity [5]. We present the detailed analyses of polar-angle-dependent optical response and near-field phenomena observed for an NbN pattern acting as S-structure.

## 2. Model system

We simulated devices consisting of cavity-based structures similar to those described in reference [2]. A 200-nm-period NbN pattern with 50 % filling factor consisting of 4-nm-thick NbN stripes, and a 2-nm-thick NbNO$_x$ dielectric cover layer is the conventional starting structure in SNSPD designs. These NbN patterns were arrayed below an unpatterned 279-nm-thick hydrogen-silsesquioxane (HSQ) dielectric film forming a quarter-wavelength optical cavity, and closed by a 60-nm-thick gold film acting as a reflector (Fig. 1a). These devices were illuminated from the bottom (sapphire substrate), side by p-polarized $\lambda$= 1550 nm light.

Figure 1b shows the geometry of the off-axis illumination in a coordinate system, where the *x-y* plane is parallel to the substrate surface. The angle $\varphi$ indicates the polar angle relative to the normal vector of the substrate surface, and $\gamma$ refers to the azimuthal angle between the plane of light incidence and the nanowire's long axes. Figure 1c indicates the p-polarized light illumination of P- and S-structures.

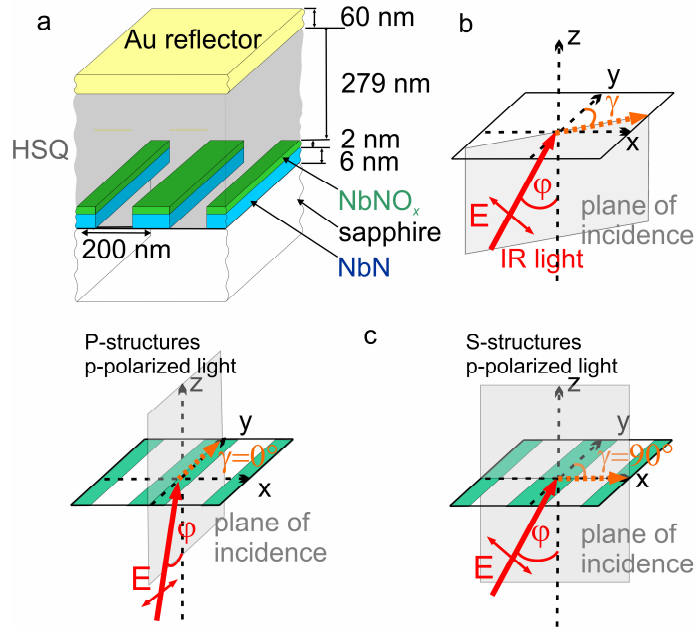

**Fig. 1** (a) Schematic drawing of structures studied: NbN patterns with 200 nm periodicity, 4 nm thickness and 50 % filling factor, covered by 2 nm NbNO$_x$ layer, are arrayed below HSQ-filled nano-optical cavities having 279 nm thickness, and covered by continuous gold film with 60 nm thickness. (b) The coordinate system as defined for the device in (a), indicating how the integrated structures are illuminated in conical-mounting by p-polarized $\lambda = 1550 \, \text{nm}$ light from the bottom (sapphire substrate) side. The $\varphi$ polar angle is measured relative to the surface normal, while the azimuthal orientation is specified by the $\gamma$ angle between the plane of light incidence and the long axis of the NbN stripes. (c) The relative orientation of the plane of incidence with respect to P-($\gamma = 0°$) and S-structures ($\gamma = 90°$).

The EM-field was specified by the **H**-field of the plane-wave with in-plane polarization, taking the $\varphi$ polar and $\gamma$ azimuthal angles into account in the specification of the three **H**-field vector components, corresponding to *x*, *y*, and *z* directions:

$$H_x \exp\{-j[(k_x x)+(k_y y)+(k_z z)]\}, \text{ where } H_x = H_o \cos\gamma, \quad (1a)$$
$$H_y \exp\{-j[(k_x x)+(k_y y)+(k_z z)]\}, \text{ where } H_y = -H_o \sin\gamma, \quad (1b)$$
$$H_z \exp\{-j[(k_x x)+(k_y y)+(k_z z)]\}, \text{ with } H_z = 0; \quad (1c)$$

while the components of the **k** wave vector were as follows:

$$k_x = k_o (\sin\varphi \sin\gamma), \quad (2a)$$
$$k_y = k_o (\sin\varphi \cos\gamma), \quad (2b)$$
$$k_z = k_o \cos\varphi. \quad (2c)$$

In the case of non-perpendicular incidence and arbitrary azimuthal angle, all components of the electric field are non-zero (Fig 1b).

An infrared light beam with $\lambda = 1550$ nm wavelength, having power of $P = 2 \times 10^{-3}$ W was used in the calculation for illumination of the NbN pattern from the substrate side [10].

## 3. Theoretical method

The calculation methods used for simulating off-axis illumination with arbitrary polarization are somewhat more complex than those used for normal-incidence illumination. For example, a simple two-dimensional model is no longer sufficient to describe the system, instead a three-dimensional model must be adopted and appropriate boundary conditions must be chosen.

Because of the possibility of inherent numerical instabilities in this new approach, we used TMM calculations to verify the FEM results. However the TMM results themselves provided useful insights into the optical physics of the device. These calculations allowed us to better understand how the sub-wavelength-thickness lossy metal layers in this cavity-based structure frustrated the total-internal reflection (relative to TIR at interface of semi-infinite bounding dielectric media). This effect is well-known in the optics thin metal films, and is referred to as attenuated TIR [10].

### *3.1. Three-dimensional numerical optical model*

Three-dimensional FEM models were prepared by using the RF module of the COMSOL software package comprising the full-wave solutions of Maxwell equations. These 3D FEM models can determine the effect of polar and azimuthal illumination angles on the optical response of the cavity-integrated structures, and can map the EM near-field distribution around the absorbing NbN-segments inside these structures.

The illumination was defined by using a port boundary condition applying boundary pairs, which makes the simultaneous variation of the polar and azimuthal angles possible, and also permits polar angles larger than the angles of TIR.

Floquet periodic boundary conditions were applied at the parallel vertical sides of the unit cells to reflect the periodic nature of the problem. Perfectly matched layers (PML) were introduced at the upper and lower boundaries to minimize spurious reflections. The upper and lower boundaries were placed $\lambda/n$ above the gold and below the sapphire, respectively. Periodic meshes were required because of the periodic nature of the problem, and the introduction of the port boundary pairs also made assembly meshing necessary.

The absorptance was determined in the NbN and Au by dividing the resistive (Joule) heating calculated in the regions containing these metals, by the incident power. The reflectance and transmittance were calculated by dividing the power flows out the bottom and top boundaries of the PMLs by the incident power.

Illumination parameters were varied to realize angle-dependent calculations with increasing angular resolutions. First we performed a coarse dual-angle-dependent study, by which we mean that both the polar and azimuthal angles were varied within the range $\varphi = [0°, 85°]$ - $\gamma = [0°, 90°]$ with angular resolution $\Delta\varphi = \Delta\gamma = 5°$ in order to provide an overview of the impact on absorptance of illumination conditions. The result of this study is presented in the graph in Figure 2.

The two cases, where methods relying on effective medium theory cannot be applied are i) the p-polarized illumination of S-structures, and ii) s-polarized illumination of P-structures. We have selected the first one for detailed studies, since p-polarized illumination is better for use with noble metal antennas for SNSPDs [11], a topic of recent interest. The special case of p-polarized illumination of the integrated pattern acting as an S-structure ($\gamma = 90°$) was investigated with high resolution by varying the polar angle in $\Delta\varphi = 0.005°$ steps in the $\varphi = [34.5°, 35.5°]$ interval, then $\Delta\varphi = 0.05°$ resolution was used between 34.0° and 34.5° and 35.5° and 36.0°. These results were combined with results of computations at lower $\Delta\varphi = 1°$ resolution across the entire $\varphi = [0°, 90°]$ polar angle region (Figure 3a and b). All of the partial solutions were stored in order to permit subsequent near-field distribution analyses (Figure 4).

### 3.2. Transfer Matrix Method computations

We performed classical TMM calculations on different unpatterned multilayered structures that occur within the structure shown in Figure 1a (as described in the paragraph below) in order to investigate the dominant optical phenomena at work, and to identify the corresponding characteristic extrema in the optical response [4]. In the device structure under study, the NbN layer was patterned. Vertical cross-sections through the film stack would thus either contain or not contain NbN depending on where the cross-section was taken. These two vertical thin-film stacks were: (1) NbN-$NbNO_x$ bi-layers, 279 nm HSQ filled cavity, and 60 nm thick Au cover-layer (referred to as stack 1); and (2) 285 nm HSQ layer between sapphire-substrate and Au-reflector (referred to as stack 2). TMM calculations were realized by approximating the cavity-based structure by an artificial composite made of these film stacks, following the approach of reference [1]. The absorptance, reflectance, and transmittance were determined by first calculating the responses of unpatterned films with the composition of each of the stacks, and then forming a weighted sum of the results, weighting each stack by its corresponding 50% fill-factor.

## 4. Results and Discussion

The most important result of these calculations was that optimization of the illumination angle and orientation of the absorbing patterns can help to optimize device absorptance, e.g. in SNSPD applications (Fig. 2). Based on FEM results, it is possible to select the optimal relative orientation of the NbN stripes with respect to the incidence plane. Furthermore, these studies serve to illuminate the basic optics at play. We present here in detail results for a study of p-polarized light illumination of S-structures.

### *4.1. Optical response*

The numerical computation, whose results are shown in Figure 2, demonstrated that the off-axis p-polarized illumination of niobium-nitride arrays resulted in a variation of NbN absorptance with polar angle that was similar across all azimuthal angles. By contrast, the variation with azimuthal angle was less noticeable.

Figure 2 shows that the absorptance was larger for p-polarized illumination of P-structures ($\gamma = 0°$) i.e. when the projection of the **E**-field in the surface plane oscillated parallel to the stripes, than it was for p-polarized illumination of S-structures ($\gamma = 90°$) i.e. when the projection of the **E**-field in the surface plane oscillated perpendicular to the stripes. Notice that this projected **E**-field will vary with polar angle. The figure also indicates a global minimum similar to previous results in the literature [3, 5]. This minimum was followed by a local maximum across all azimuthal angles.

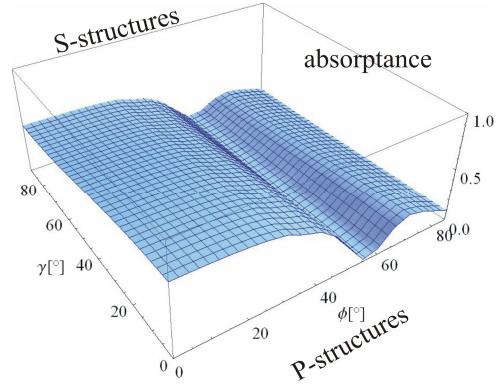

**Fig. 2** The dual-angle-dependent absorptance of the 200 nm period NbN stripes embedded in an optical cavity. The calculations were performed over the $\gamma = [0-90°]$ and $\varphi = [0-85°]$ intervals, with $\Delta\gamma = \Delta\varphi = 5°$ resolution.

Figure 3a and 3b show in more detail the effect of polar angle variation for p-polarized illumination of the NbN pattern acting as an S-structure ($\gamma = 90°$). In Figure 3a, we see 46.5 % NbN absorptance at perpendicular incidence, which decreases to nearly zero absorptance at $\varphi = 57°$. The NbN absorptance remains smaller than at perpendicular incidence throughout all polar angles, but reaches a local maximum at $\varphi = 70°$, where the value is approximately a third of the absorptance at perpendicular incidence.

The absorptance of the bare NbN stripes was calculated by using FEM analysis on the same pattern, also illuminated from below. At normal incidence, this structure (without the cavity) had an absorptance of 25%. For p-polarized illumination of S-structures, the cavity enhances the absorption efficiency relative to a bare film. This was shown to be true up to polar angles ($\varphi$) of approximately 47°.

We also observed a local NbN absorptance minimum, a resonant dip in reflectance and a coincident enhancement in gold absorptance in a narrow polar-angle region between 34.5° and 35.5°. Figure 3b shows high-resolution computations that analyze this region in more detail. Furthermore, to ensure that the fine structure of the optical response was not simply the result of numerical artifacts, we compared FEM results to corresponding curves determined by TMM computation. This comparison revealed that analogous extrema appear in the TMM calculations in the total absorptance and also in the reflectance. The only qualitatively significant difference between the FEM and TMM approach was in the transmittance signals (Fig. 3b). We will discuss this difference briefly below.

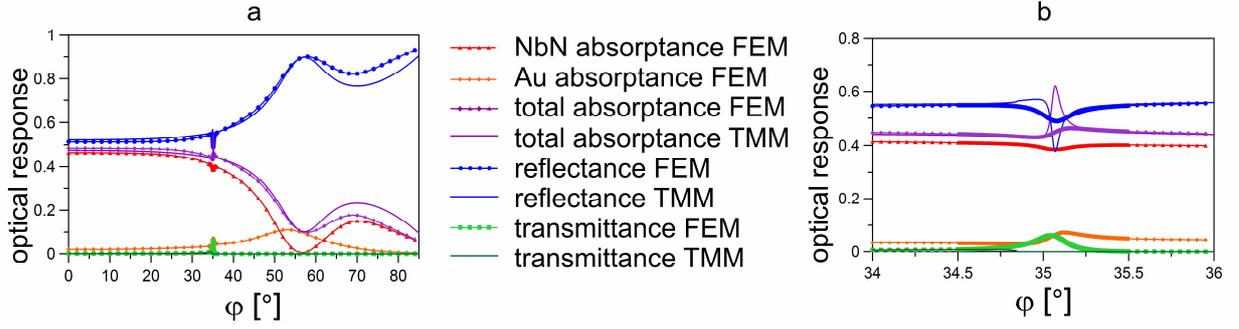

**Fig. 3** (a-b) The comparison of the optical responses at $\gamma = 90°$ determined by FEM (lines with symbols) and by TMM (lines). (b) The results of FEM calculations in [34.5°, 35.5°] interval with $\Delta\varphi = 0.005°$ resolution are incorporated into the graph prepared in [34°, 36°] interval with $\Delta\varphi = 0.05°$ resolution, and all of these data are incorporated into (a) graphs originating from computation performed with $\Delta\varphi = 1°$ resolution in [0°, 85°] region. The "absorptance – reflectance - transmittance TMM" refers to the optical response of composites made of NbN-NbNO$_x$-HSQ-Au and HSQ-Au film stacks, according to the 50 % fill factor in 200 nm periodic cavity-based structure. There is a local and global minimum on the NbN absorptance at 35° and 57°, respectively. The origins of these minima are detailed in the text.

The absorptance is larger when the **E**-field is oriented parallel to the NbN stripes (as is the case for p-polarized illumination of P-structures), because **E**-field oscillation parallel to the wires promotes field penetration [12]. The TMM used weighted combinations of optical responses on unpatterned films, therefore it could not account for this penetration effect. FEM can include optical effects occurring due to the pattern geometry itself (rather than just the fill factor), while TMM cannot. Thus, a more pronounced difference was observed between optical responses of P-structures calculated using FEM and TMM.

On S-structures the difference manifested itself in slightly higher FEM NbN absorptance at small polar angles. The difference was more pronounced and the FEM absorptance was smaller at large polar angles, e.g. around the local maximum of absorptance at 70°.

The global minimum in NbN absorptance, occurring at 57°, is beyond the angle of total internal reflection of a semi-infinite sapphire-HSQ interface, which is 52.6°. This effect is understood to be due to the presence of the gold and NbN near the sapphire-HSQ interface. Figure 3a shows a broad global maximum in gold absorptance at 53° calculated by FEM. The presence of a maximum in gold absorptance near this angle suggests that the minimum in NbN absorptance at 57° is in the field of attenuated TIR caused by the presence of gold.

Fig. 3 shows that the local minimum in NbN absorptance at 35° appears slightly beyond the angle corresponding to TIR phenomenon occurring at 34.85° for a bare NbN layer on a sapphire substrate [5], and is accompanied by a reflectance dip. Figure 3 also indicates that the local maximum in gold absorptance is at  polar angle, exactly coincident with the local NbN absorptance and reflectance minima. This coincidence is due to a plasmonic mode excitation in the gold that competes with both the NbN absorptance and reflectance at this orientation. The surface plasmon polariton is excited at an angle of 35.1°.

These observations confirm that neither global nor local NbN absorptance minima could originate from simple TIR-related phenomenon, as was the case in reference [5]. Instead, both minima occur in connection with phenomena accompanying reflection from and absorption in the gold layer. The identification of their origin requires the detailed investigation of the near-field at these specific orientations.

The FEM optical responses on S-structures are in quantitatively good agreement with weighted TMM responses of film stacks at small polar angles, indicating that the local extrema might be identified by the TMM. The difference in FWHM of extrema on FEM and TMM curves is not caused by the difference in computation resolutions, since both were 0.005° in the $\varphi = [34.5°, 35.5°]$ interval. Differences appearing in transmittance might be explained by the fact that the signal was collected at a plane in the reflector's near-field to determine transmittance in FEM, therefore it contains information about near-field phenomena, which are outside the scope of TMM.

Because the model is defined on a finite vertical region, all optical responses determined by FEM are technically near-field responses. This explains, for example, why some transmittance is observed in the FEM calculations above the angle of total-internal reflection of sapphire and air. However, these errors are small, and in general we are most interested in the near-field interactions between device elements.

*4.2. Near-field phenomena*

The detailed investigation of the normalized **E**-field along cross-sections taken perpendicular to the niobium stripes shown in Fig. 4a indicates dependence of the nanometer-length-scale variation of amplitude and phase around the NbN segments on the polar-angle of the illuminating field (Fig. 4b).

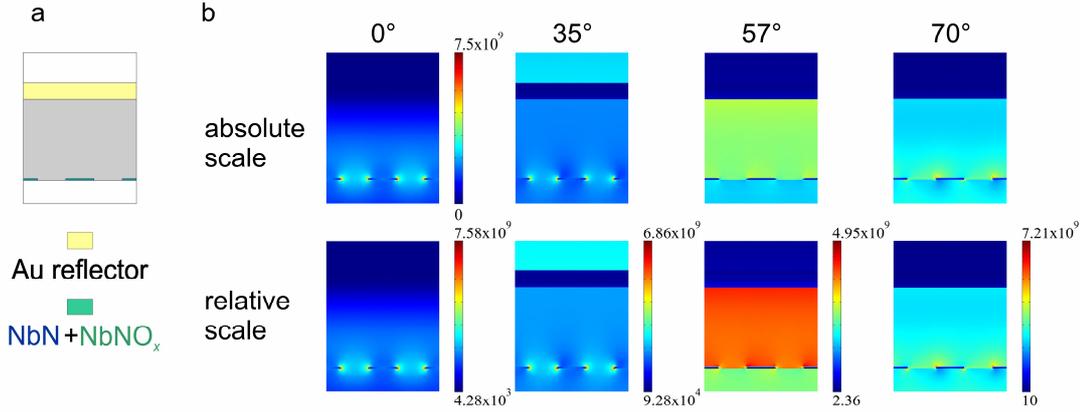

**Fig. 4** (a) Schematic drawing showing two unit cells in a plane perpendicular to the NbN stripes. The pattern period is 200 nm. (b) The **E**-field distribution in presence of nano-cavities and gold reflector taken at different polar angles at the plane shown in (a). The pictures in the upper row are plotted using the same scale in [V/m], while the pictures on bottom are scaled to illustrate better small variations in the field.

At small polar angles, an **E**-field enhancement is observable around the NbN segments, which manifest itself in the highest intensity at the corners of the stripes (Fig. 4b, 0° and 35°). Surprisingly, at the polar angle of 35°, corresponding to the most sudden changes in the optical response, the **E**-field is enhanced at the air side of the reflector (Fig. 4b, 35°), and the resistive heating (not shown in the Figure) is enhanced at *both* dielectric interfaces of the gold film. At the polar angle of 57° corresponding to the minimum in NbN absorptance, the **E**-field distribution is concentrated in the quarter-wavelength cavity, rather than around the NbN stripes (Fig. 4b, 57°), and the resistive heating map (not shown) indicates significantly increased values at the gold-HSQ interface. At 70° polar angle corresponding to the local NbN absorptance maximum a very intense and anti-symmetric **E**-field distribution was observed around the NbN stripes, i.e. those NbN corners are more bright, where the oblique incident beam first reaches them (Fig. 4b, 70°).

As a result of the ~ quarter wavelength cavity, the NbN segments are in the area of **E**-field antinodes at small incidence angles, which results in absorptance enhancement compared to bare NbN patterns presented in previous studies (Fig. 4b, 0° and 35°) [5]. This benefit originating from cavity integration was confirmed experimentally previously [3]. The field redistribution observed at 57° polar angle in the cavity reveals that the waves incident on NbN from the substrate and back-reflected from the gold are out-of phase, which cause negligible remaining EM-field intensity that might be absorbed inside the NbN segments. The large reflectance observed at this orientation suggests that there is phase-matching between the waves reflected from NbN and from Au. An **E**-field enhancement is observable at the top of the cavity below the gold reflector, and the resistive heating map (not shown) reveals that this near-field distribution is accompanied by absorptance in gold, corresponding to the attenuated TIR characteristics of the reflectance curves at this orientation.

At the same time, a surface plasmon mode confined along the Au-HSQ boundary appears at this orientation (Fig. 4b, 57°). This result shows that the presence of filling material in the quarter wavelength cavity with refractive index smaller than that of sapphire makes it possible to excite and guide surface modes on the bottom side of the reflector. Excitation of such a mode could be beneficial in other applications.

FEM computation results indicate that is possible to avoid losses due to reflection and to absorption in the gold, by further increasing the polar angle. The final absorptance is limited by the polar angle dependent light in-coupling efficiency into multipolar **E**-field around the tiny NbN stripes. At e.g. 70° polar angle in which case approximately 15 % absorptance might be reached in NbN segments, the spatial distribution of the instantaneous **E**-field-components indicates more pronounced quadrupolar characteristics compared to that observable in the angular region around the global minimum.

The near-field cross-section at polar angles near 35° reveals the propagating plasmon modes at the air side of the structure, implying that the EM-field has passed through the gold reflector at this angle. The possibility of field-enhancement at the air side of thin metal films is a known phenomenon, which is in connection with surface plasmon polariton excitation [13]. **E**-field enhancement above the reflector occurs at this angle, and the appearance of the local minimum on NbN absorptance might be explained by a portion of the **E**-field being out-coupled from the cavity (Fig. 3b and 4b, 35°). The increased resistive heating on the top of the reflector caused by presence of these surface modes corresponds to the local maximum of gold absorptance at 35°. The decay length of these modes in air is larger than the distance to the upper PML, where the transmitted signal was collected, therefore their presence causes apparent transmittance. In a far-field experiment, however, this transmittance could not be observed.

Further enhancement of the coupling might be achieved at a specific illumination orientation or at a variety of orientations by controlling the local EM-field distribution around the NbN segments via appropriately designed Au cavity-structures, which were previously studied in case of perpendicular light incidence [14].

## 5. Conclusions

The key result of this work is our demonstration that three-dimensional FEM calculations can be used to optimize the illumination angle in nanowire detector structures. The particular cavity-based structure we choose to study most closely did exhibit enhanced absorptance relative to bare NbN patterns, and the method that we developed here is broadly applicable. Along the way, we have shown interesting optical physics involving the cavity and the overlying gold film that occurs under precisely defined illumination conditions. The advantage of the developed numerical method is its ability to describe these complex effects.

Further studies could repeat these calculations, but using integrated noble-metal antenna arrays capable of optimizing different optical and resonance phenomena simultaneously. This method could also optimize the absorptance of an NbN pattern acting as S- and P-structure when illuminated by p-polarized light.

**Acknowledgement**

This work has been supported by the U.S. Dept. of Energy Frontier Research Centers program. M. Csete would like to thank the Balassi Institute for the Hungarian Eötvös post-doctoral fellowship. The authors would like to thank for the helpful discussions with E. Driessen and M. de Dood, and COMSOL engineers in Burlington. Áron Sipos would like to thank for the support of OTKA foundation from the National Office for Research and Technology Developments (NKTH), grant No OTKA-NKTH CNK 78459.

**References**

[1] G. N. Gol'tsman, O. Okunev, G. Chulkova, A. Lipatov, A. Semenov, K. Smirnov, B. M. Voronov, A. Dzardanov, C. Williams, and R. Sobolewski, „Picosecond superconducting single-photon optical detector," Appl. Phys. Lett. **79/6** 705-708 (2001).

[2] K. M. Rosfjord, J. K. W. Yang, E. A. Dauler, A. J. Kerman, V. Anant, B. M. Voronov, G. N. Gol'tsman, and K. K. Berggren,"Nanowire Single-Photon detector with an integrated optical cavity and anti-reflection coating," Optics Express **14/2** 527-534 (2006).

[3] V. Anant, A. J. Kermann, E. A. DAuler, J. K. W. Yang, K. M. Rosfjord, K. K. Berggren, „Optical properties of superconducting nanowire single-photon detectors," Optics Express **16/14** 10750-10761 (2008).

[4] M. Born, E. Wolf, Principles of Optics (Pergamon Press, 1964).

[5] E. F. C. Driessen and M. J. A. de Dood,"The perfect absorber," Appl. Phys. Lett. **94** 171109/1-3 (2009).

[6] M. G. Moharam, T. K. Gaylord, "Rigorous coupled-wave analysis of planar-grating diffraction," J. Opt. Soc. Am. **71/7** 811-818 (1981).

[7] X. J. Yu and H. S. Kwok, „Optical wire-grid polarizers at oblique angles of incidence" Journal of Appl. Phys. **93/8** 4407-4412 (2003).

[8] E. F. C. Driessen, F. R. Braakman, E. M. Reiger, S. N. Dorenbos, V. Zviller, M. J. A. de Dood, „Impedance model for the polarization-dependent optical absorption of superconducting single-photon detectors," The European Journal Applied Physics **47** 1071/1-6 (2009).

[9] L. Li, „A modal analysis of lamellar diffraction gratings in conical mountings," J. Mod. Opt. **40/4** 553-573 (1993).

[10] Saleh-Teich, Fundamentals of Photonics (Wiley, 2007):

The relationship between the intensity and **E**-field magnitude is:

$$I = \frac{|E_o|^2}{2\eta}, \quad (3)$$

where the impedance $\eta$ is defined in terms of free space impedance $\eta_o$ via $\eta = \eta_o/n_{\text{sapphire}}$, and taking into account that our substrate is a non-magnetic medium: $\mu = \mu_o$, whereupon $\eta = \sqrt{\mu_o/\varepsilon}$; also, sapphire is essentially non-absorbing at $\lambda = 1550\,\text{nm}$, therefore $\sqrt{\varepsilon} = n_{\text{sapphire}}$, so the amplitude of the **E**-field was determined as follows:

$$E_\text{o} = \sqrt{2I\eta} = \sqrt{2\frac{P}{A}\frac{1}{n_{sapphire}}}\sqrt{\frac{\mu_\text{o}}{\varepsilon_\text{o}}} \ . \qquad (4)$$

Based on this calculation the power-flow at the source boundary corresponds to $E_\text{o} = 2.32\times10^6$ V/m **E**-field amplitude. This value for $E_\text{o}$ is the reference amplitude necessary to interpret the near-field distribution determined by COMSOL models.

[11] X. Hu, C. W. Holzwarth, D. Masciarelli, E. A. Dauler, and K. K. Berggren, „Efficiently Coupling Light to Superconducting Nanowire Single-Photon Detectors," IEEE Transactions on Applied Superconductivity **19/3** 336-340 (2009).

[12] G. R. Bird, M. Parrish,"The Wire Grid as a Near-Infrared Polarizer," Journal of Optical Society America **50/9** 886 -891(1960).

[13] D. K. Gramotnev, S. I. Bozhevolnyi, "Plasmonics beyond the diffraction limit," Nature Photonics **4** 83-91 (2010).

[14] X. Hu, E. A. Dauler, R. J. Molnar, K. K. Berggren, „Superconducting nanowire single-photon detectors integrated with optical nano-antennae," Optics Express **19/1** 17-31 (2010).